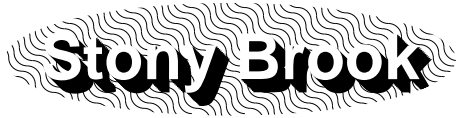

# THE BIG PICTURE

N. Berkovits, M.T. Hatsuda, and W. Siegel[1]

*Institute for Theoretical Physics*
*State University of New York, Stony Brook, NY 11794-3840*

**ABSTRACT**

We discuss the conformal field theory and string field theory of the NSR superstring using a BRST operator with a nonminimal term, which allows all bosonic ghost modes to be paired into creation and annihilation operators. Vertex operators for the Neveu-Schwarz and Ramond sectors have the same ghost number, as do string fields. The kinetic and interaction terms are the same for Neveu-Schwarz as for Ramond string fields, so spacetime supersymmetry is closer to being manifest. The kinetic terms and supersymmetry don't mix levels, simplifying component analysis and gauge fixing.

---

[1] Work supported by National Science Foundation grant PHY 89-08495.

Internet addresses: nathan, mhatsuda, and siegel@max.physics.sunysb.edu.

# 1. INTRODUCTION

## 1.1. Overview

In the Neveu-Schwarz-Ramond treatment of the superstring, the bosonic ghost of local world-sheet supersymmetry is necessary for the construction of spinor vertex operators and space-time supersymmetry. Unfortunately, the zero-mode of this ghost causes an ambiguity in the cohomology of the Becchi-Rouet-Stora-Tyutin operator. Fixing this ambiguity by choice of a "picture" leads to improper ghost number assignments, which must be fixed by the insertion of picture-changing operators. In the resulting string field theory these picture-changing operators appear in the kinetic terms in a way which mixes levels, preventing a component analysis of the gauge-invariant lagrangian. The OSp(1|2)-invariant "vacuum" is not annihilated by one of the negative-"energy" ((mass)$^2$) bosonic ghost oscillators ($\gamma_{1/2}$), so ghost states of arbitrarily negative energy can be created.

In a previous paper [1] such complications were avoided by a general method which can be applied to any BRST formulation with unpaired, unphysical bosonic modes. The addition of a nonminimal term to the BRST operator, corresponding to the choice of a Lorentz gauge (any gauge involving one time derivative on the gauge field), introduces an additional bosonic mode which can be paired with the other to form creation and annihilation operators, defining the vacuum unambiguously. After this minor change, the BRST analysis is standard: The BRST cohomology is unique, the physical states have the right ghost number, and $b_0 = 0$ is sufficient to fix the gauge.

In this paper we construct NSR conformal field theory and string field theory along the lines outlined in [1]. The first-quantization corresponds to gauges with a propagating world-sheet gravitino. All (unintegrated) vertex operators in the conformal field theory have ghost number one. In particular, even though the NS Hilbert space is essentially the same (up to nonminimal fields) as in the usual treatment, the vertex operators are unique, are closely analogous to those in the Veneziano string, and differ from the usual choice because of a new unique choice of "vacuum." We show the equivalence of general Neveu-Schwarz N-point tree amplitudes to the usual results by applying conformal field theory to our formulation in an explicit calculation. We also discuss amplitudes involving massless spinors.

As in NSR light-cone string field theory, the treatment of the NS and R string fields is now identical, except for the usual difference in boundary conditions on the world sheet: Both string fields are fermionic (the tachyon "ground state" is an unphysical fermion), and have $\Phi^\dagger Q \Phi$ kinetic terms without "picture-changing" insertions, allowing component analysis and simple gauge fixing. The (NS)$^3$ vertex is essentially the same as in Witten's original version of NSR superstring field theory [2], except for the nonminimal coordinates. However, unlike that original version, the NS(R)$^2$ vertex is now the same as the (NS)$^3$ one, except that again the boundary conditions differ. The same supersymmetry operator transforms NS to R fields as R to NS.

## 1.2. Outline

In the following section we describe the relation between BRST first-quantization with vertex operators and Zinn-Justin-Batalin-Vilkovisky second-quantization. Both methods can be applied to particles as well as strings. In particular, in Yang-Mills theory there is a "vacuum" state from which the physical states can be obtained by applying the vertex operator, just as in string theory.

In section 3 we consider the relation between these two methods in more detail for particles: Yang-Mills and the Dirac spinor. In particular we examine the nonminimal fields which will be used in the massless sector of the NSR string later. We show how this allows the choice of harmonic oscillator boundary conditions, which provide a unique picture for the cohomology, so that the free Dirac spinor action does not require picture-changing insertions. The fields obtained for this super-Yang-Mills system (Yang-Mills+spinor) resemble those found in the manifestly supersymmetric version of 4D N=1 super-Yang-Mills. The operator formalism for interacting Yang-Mills is also described.

The first-quantization of the free NSR string with nonminimal coordinates is described in section 4. By starting with the hamiltonian formalism the desired result can be obtained without field redefinitions and with little modification from the method used for the Dirac spinor. The conformal weights for the nonminimal coordinates are arbitrary, but match between the fermions and the bosons, guaranteeing conformal anomaly cancellation. (Similar remarks apply to choices of weights for ghosts in the Green-Schwarz string.)

Interactions for the first-quantized theory are treated in section 5 by the method of conformal vertex operators. We describe some general relations between the integrated and unintegrated forms of these operators which apply for our generalizations of the conformal gauge. In the Neveu-Schwarz sector, the vertex operators can be written without the use of bosonization. The physical ones are independent of the nonminimal coordinates, and so can also be written in world-sheet superspace. To treat Ramond vertex operators, and to relate our picture to the usual OSp picture, we give the form of bosonization corresponding to our harmonic oscillator boundary conditions.

In section 6 we give the supersymmetry operator. Since in our picture all unintegrated (integrated) vertices have ghost number 1 (0), so all string wave functions/fields have ghost number $-\frac{1}{2}$, the same supersymmetry operator can be used to transform from NS→R and R→NS. This is possible because, as in manifestly spacetime supersymmetric formulations of particle theories, the supersymmetry generator contains the analog of both the "$\partial/\partial\theta$" and "$\not{p}\theta$" terms, whereas the old formulation was more like the component approach in that these two terms appeared in the two separate transformations (i.e., in two different pictures). As in the old formalism, the supersymmetry transformation is linear in the string fields.

We discuss scattering amplitudes in section 7, and calculate the general NS amplitude (explicitly evaluating the ghost matrix elements). In practice, the calculations do not differ much from the Friedan-Martinec-Shenker-Knizhnik treatment. The main conceptual difference is that the FMSK approach required two pictures for any type of vertex operator, as determined by ghost number conservation (modulo the anomaly). On the other hand, in our formalism the vertex operator is always in the same picture but contains terms similar to those appearing in both pictures of the FMSK approach (as does the supersymmetry operator, discussed above). However, only one type of term is chosen from any particular vertex operator because of the way the ghosts appear. Although NS amplitudes are slightly simpler than in the usual formulation, amplitudes involving spinors are somewhat messier in their ghost dependence, and we just set up the formalism. (However, the main problems in both formalisms are evaluating matrix elements of physical spin operators and deriving massive fermion vertex operators.)

In section 8 we discuss the superstring field theory. The NS and R string fields appear identically, except for the usual integer vs. half-integer mode numbers. Bosonization is unnecessary, since the NS and R Hilbert spaces can be treated as independent. Both string fields are fermionic (but GSO picks out the right parts), both have the BRST operator as kinetic operator without picture-changing insertions, both use the same supersymmetry operator (as discussed earlier), and both interaction terms $(NS(R)^2$ and $(NS)^3)$ have the same operator insertion $\psi \cdot p + \cdots$. (Picture-changing exponentials, which appear in the super-Riemann surface formulation, cancel when one uses unbosonized ghosts and vacuua in the physical picture.) This suggests that this form is one step closer than the old formalism to a covariantly quantized Green-Schwarz formalism.

In the final section we state our conclusions, including a few conjectures based on some of the more detailed results discussed below.

## 2. SECOND QUANTIZATION VIA FIRST

### 2.1. BRST cohomology

In general, first-quantized relativistic systems can be described completely by a (hamiltonian) BRST operator $Q$ and a ghost-number operator $J$. $Q$ is of the form (ghost)×(constraint) + (ghost terms), where the ghosts and constraints are unphysical operators. Any of these operators which could have continuous eigenvalues must be paired up as creation and annihilation operators, except for the constraint $p^2 + M^2$. ($p^2 + M^2$ is treated differently because it is not a true constraint: In the classical mechanics lagrangian, its Lagrange multiplier is restricted to be positive, so the propagator is $\sim 1/(p^2 + M^2)$, not $\delta(p^2 + M^2)$.) This may require the introduction of extra variables as nonminimal terms of the form $ab$, where $a$ is a commuting c-number and $b$ anticommuting, so $a$ can be paired with the unpaired boson that appears in the minimal part of the BRST operator [1].

The physics of the system is then described by the cohomology of $Q$: states $|\psi\rangle$ which satisfy the equation of motion $Q|\psi\rangle = 0$ modulo the gauge transformation $\delta|\psi\rangle = Q|\lambda\rangle$. These states can be classified by their ghost number, which can take four values: $-\frac{1}{2}$ for the physical states, $-\frac{3}{2}$ for the "physical gauge parameters," $+\frac{1}{2}$ for the antifields of the physical states, and $+\frac{3}{2}$ for the antifields of the physical gauge parameters. (For the Veneziano string, see [3]. In general, there could be zero-momentum states in the cohomology at $J$ other than $\pm\frac{3}{2}$, but not for the theories we consider, which have only Yang-Mills as massless gauge fields.) The physical gauge parameters have ghost number one less than the physical states because $Q$ has ghost number one $(\delta|\psi\rangle = Q|\lambda\rangle)$, and "physical" means that they are the global symmetries which survive gauge fixing [4]: In the field theory the gauge transformation has the form $\delta\psi = Q\lambda + \psi * \lambda - \lambda * \psi$ (+ perhaps higher-order terms in $\psi$), where the form of "*" depends on how $\psi$ represents the global group. When $Q\lambda = 0$, there is no inhomogeneous piece, so that $\lambda$ does not gauge away part of some gauge field. On the other hand, when $\lambda = Qf$ for some $f$, then there is another gauge parameter $\lambda' = \psi * f - f * \psi$ (+ higher-order terms in $\psi$) that produces that same transformation (up to transformations of the form $\delta\psi_i = \epsilon_{ij}\delta S/\delta\psi_j$ for graded antisymmetric $\epsilon$, which are invariances of any action). Thus the nontrivial global parts of the gauge symmetries are those in the BRST cohomology. (For example, the parts of the vector ghost in the cohomology for gravity correspond to global Poincaré transformations, while the global parts of antisymmetric tensor gauge transformations do not appear in the cohomology.)

## 2.2. ZJBV quantization

The antifields are the same ones that appear in Zinn-Justin-Batalin-Vilkovisky quantization [5]: The second-quantized formalism which follows from BRST first-quantization [2] is equivalent to ZJBV second-quantization [6]. Specifically, if we expand a general state as $|\psi\rangle = |\psi_+\rangle + |\psi_-\rangle$, where $b_0|\psi_+\rangle = c_0|\psi_-\rangle = 0$, $c_0$ is the ghost for $p^2 + M^2$, and $b_0$ is its canonical conjugate ($\{b_0, c_0\} = 1$), then $|\psi_+\rangle$ corresponds to a field while $|\psi_-\rangle$ corresponds to an antifield. The Hilbert-space inner product is nonvanishing only between fields and antifields, since it includes integration over $c_0$, and this inner product is therefore fermionic, and corresponds to the ZJBV "antibracket" ( , ): Writing the field as a state $|\Phi\rangle$,

$$(\langle A|\Phi\rangle, \langle \Phi|B\rangle) = \langle A|B\rangle \tag{2.1}$$

represents the obvious bracket between two $|\Phi\rangle$'s, but with the unusual property of being fermionic because of the anticommutativity of integration over $c_0$. (As usual, the Hilbert-space inner product $\langle A|B\rangle$ can be written in a coordinate representation as $\int A^\dagger B$.) Any state and its "dual" antistate (relative to the inner product: for $\langle \psi_{+i}|\psi_-^j\rangle = \delta_i^j$, $|\psi_-^i\rangle$ is the antifield to $|\psi_{+i}\rangle$) have opposite ghost number ($J$ is antihermitian) and opposite statistics. The "$S$" operator of the ZJBV formalism is

$$S = \tfrac{1}{2}\langle\Phi|Q\Phi\rangle + \tfrac{1}{3}g\langle\Phi|\Phi \star \Phi\rangle \tag{2.2a}$$

(+ perhaps higher-order terms), and both contains the equations of motion and generates the BRST transformations (as does $Q$ for the corresponding free first-quantized theory). $S$ is bosonic, as follows from the fermionic nature of both $Q$ and the inner product. It can also be written as

$$S = \tfrac{1}{2}\langle\Phi|Q\Phi\rangle + \tfrac{1}{3}g\langle V_3|\Phi\rangle|\Phi\rangle|\Phi\rangle \quad, \tag{2.2b}$$

where $|V_3\rangle$ is some three-particle(/string) state. The analog of $Q^2 = 0$ is

$$(S, S) = 0 \quad, \tag{2.3}$$

which by a perturbation in $g$ implies: (1) $Q^2 = 0$, (2) $Q$ is distributive over the $\star$ product (i.e., $\star$ and $|V_3\rangle$ are $Q$-invariant), and (3) the $\star$ product is associative (up to higher-order terms). It also implies

$$\delta\Phi = (A, \Phi) \quad, \quad A = (S, B) \quad \Rightarrow \quad \delta S = 0 \tag{2.4a}$$

for any $B$, since then $\delta S = ((S, B), S) \sim ((S, S), B) = 0$. This generalizes to finite transformations as

$$f[\Phi] \to e^{L_A} f e^{-L_A} \quad, \tag{2.4b}$$

where $L_\alpha \beta \equiv (\alpha, \beta)$ is the Lie antiderivative. $A$ is thus the generator for gauge transformations: The usual field-independent gauge parameter $\Lambda$ appears as

$$B = \langle\Lambda|\Phi\rangle \quad \Rightarrow \quad \delta\Phi = Q\Lambda + g(\Phi \star \Lambda - \Lambda \star \Phi) \quad. \tag{2.4c}$$

For the interacting theory to be nontrivial, we must also have

$$S \neq e^{L_\Psi} \tfrac{1}{2}\langle\Phi|Q\Phi\rangle e^{-L_\Psi} \tag{2.5}$$

for any fermion $\Psi$ cubic (and perhaps higher order) in $\Phi$. This implies that $|V_3\rangle$ is also not $Q$ on something, so it is in the BRST cohomology for three-particle(/string) states.

In the first-quantized formalism, the Fermi-Feynman gauge is fixed by choosing as hamiltonian

$$H_{FF} = \{Q, b_0\} = \tfrac{1}{2}(p^2 + M^2) \quad. \tag{2.6}$$

More general gauges can be chosen by generalizing the gauge-fixing fermion $b_0$:

$$H = \{Q, b_0 + f_0\} \quad, \tag{2.7a}$$

where $f_0$ is independent of $b_0$ and $c_0$ since the $b_0$ term already takes care of fixing that one gauge invariance, or equivalently

$$H' = U^{-1}HU = \{Q', b_0\} \quad, \quad Q' = UQU^{-1} \quad, \quad U = e^{c_0 f_0} \quad, \tag{2.7b}$$

where $f_0^2 = 0$, and $H' = H$ if $[H, f_0] = 0$. The equivalent procedure in the second-quantized formalism is

$$S_{FF} = S|_{b_0 = 0} \quad, \tag{2.8}$$

where "$|_{b_0 = 0}$" means to evaluate at $b_0|\Phi\rangle = 0$ (i.e., set antifields to zero), or more generally

$$S_{gf} = S'|_{b_0=0} \quad, \quad S' = e^{L_\Psi} S e^{-L_\Psi} \quad, \quad \Psi = -\langle\Phi|c_0 f_0 \Phi\rangle \quad. \tag{2.9a}$$

Note that this $\Psi$ contains only fields, not antifields. This can be generalized: Terms in $\Psi$ linear in antifields induce field redefinitions, and terms quadratic modify the BRST transformations by terms proportional to the field equations. However, the less general case expressed in (2.9a) will generally be sufficient. It can also be written without a unitary transformation as

$$S_{gf} = S|_{b_0 + f_0 = 0} \quad. \tag{2.9b}$$

In the Feynman rules, instead of eliminating the antifields we can achieve the same effect by using $(b_0 + f_0)/H$ as the propagator.

The gauge-invariant action is obtained from $S$ by the restriction $J|\Phi\rangle = -\tfrac{1}{2}|\Phi\rangle$, which picks out the physical fields, but also includes antifields with ghost number $-\tfrac{1}{2}$, which are Nakanishi-Lautrup-type auxiliary fields.

## 2.3. Vacuum and vertex operators

All the statements above in this section apply equally well to particles and strings, bosons and fermions. As an example, consider Yang-Mills theory. The operator $Q$ can be obtained by the first-quantized methods of the covariantized light-cone (as for any other free theory) [7]. The part of the cohomology at $J = -\frac{3}{2}$ is given by the gauge parameter for global internal symmetry transformations, which is represented in the wave function (field) by the zero-momentum Faddeev-Popov ghost. (In going from gauge transformation to BRST transformation, gauge parameter is replaced by ghost.) In open string theories Yang-Mills is the only massless (i.e., unbroken) gauge field, and therefore this state is also the only state in the cohomology with $J = -\frac{3}{2}$. The state at $J = +\frac{3}{2}$ is the antifield of the Faddeev-Popov ghost at zero momentum. These two states are the only states in the cohomologies of these theories which are Poincaré invariant, since the states in the cohomology satisfy the mass shell condition (and thus must be massless for $p = 0$), and these are the only massless scalars in the cohomology. In the Veneziano string theory, this $J = -\frac{3}{2}$ state is the one invariant under the Sp(2) subgroup of the worldsheet conformal group. However, in the Neveu-Schwarz-Ramond string theory, we will find that this is not the one invariant under the OSp(1|2) subgroup of superconformal transformations. In fact, our choice of boundary conditions of the wave function in the space of the nonminimal variables we have added (and their partners) eliminates the OSp(1|2)-invariant state from the Hilbert space.

We will find that, at least for string field theory, it is more important to have a state in the cohomology with the right ghost number, and the NSR state with $J = -\frac{3}{2}$ plays the same role as in Veneziano string theory, even though it is now not Sp(2) invariant. For example, the first-quantized analog of (2.2-5) is to define interactions via vertex operators by

$$Q_{int} = Q + gV \quad, \quad Q_{int}^2 = 0 \quad, \qquad (2.10a)$$

which implies similar conditions upon perturbation in $g$. In fact, such expressions can be derived from (2.2) by expanding $\Phi$ about an on-shell background and keeping just the terms quadratic in $\Phi$, $\frac{1}{2}\langle\Phi|Q_{int}\Phi\rangle$. (This gives the usual $V$'s up to conformal transformations.) In particular, $V$ must be in the operator cohomology of $Q$:

$$Q_{int}^2 = 0 \quad \Rightarrow \quad \{Q, V\} = 0 \quad,$$

$$Q_{int} \neq e^{-\lambda} Q e^{\lambda} \quad \Rightarrow \quad V \neq [Q, \lambda] \quad. \qquad (2.10b)$$

The last relation is just the statement that $V$ is not pure gauge: A unitary transformation on $Q_{int}$ is equivalent to a gauge transformation on $V$. (Consider the corresponding statement on the Yang-Mills covariant derivative, $\nabla = p + gA$.) The infinitesimal (abelian) gauge transformation is thus $\delta V = [Q, \lambda]$.

States are then defined as

$$|\psi\rangle = V|0\rangle \qquad (2.10c)$$

in terms of a vacuum which must be in the BRST cohomology with $J = -\frac{3}{2}$:

$$J|\psi\rangle = -\frac{1}{2}|\psi\rangle \quad \Rightarrow \quad J|0\rangle = -\frac{3}{2}|0\rangle \quad,$$

$$Q|\psi\rangle = 0 \quad \Rightarrow \quad Q|0\rangle = 0 \quad, \quad |\psi\rangle \neq Q|\lambda\rangle \quad \Rightarrow \quad |0\rangle \neq Q|\lambda'\rangle \quad. \qquad (2.10d)$$

Since open strings include Yang-Mills at the massless level, we can also determine the normalization of these $J = \pm\frac{3}{2}$ states: An analysis of the BRST transformations of Yang-Mills (see e.g., [7] and below) shows that in terms of the Yang-Mills ghost $|0\rangle$ at $J = -\frac{3}{2}$ we have the Nakanishi-Lautrup field (the antifield of the antighost) $c_0|0\rangle$ at $J = -\frac{1}{2}$, the antighost $Qc_0|0\rangle$ at $J = +\frac{1}{2}$, and the antifield of the ghost $c_0Qc_0|0\rangle$ at $J = +\frac{3}{2}$. From the facts that the kinetic term $\langle\Phi|Q\Phi\rangle$ includes (Nakanishi-Lautrup)$^2$ and (antighost)$\Box$(ghost) and that $\langle$field|antifield$\rangle = 1$, we see that we have the inner product

$$\langle 0|c_0 Q c_0|0\rangle \sim 1 \qquad (2.11)$$

for Yang-Mills and thus for open strings.

The gauge fixed version of (2.10) is

$$H_{int} = \{Q_{int}, b_0 + f_0\} \equiv H + g\mathcal{W} \quad,$$

$$\mathcal{W} = \{V, b_0 + f_0\} \quad. \qquad (2.12)$$

Unlike $V$, which is "gauge covariant," $\mathcal{W}$ is gauge fixed, since it depends explicitly on the choice of $f_0$ (as does $H$).

## 3. PARTICLES

### 3.1. Yang-Mills and OSp(D,2|4)

We first consider the massless Dirac spinor and Yang-Mills as warm-ups for the NSR string, and because we can illustrate some pedagogical details which for the string would be more complicated but not more enlightening.

The method of adding equal numbers of commuting and anticommuting dimensions to the light cone, enlarging the Lorentz group to an OSp group, can be used to derive first-quantized BRST operators directly for arbitrary representations of the Poincaré group, without reference to classical mechanics actions [7]. Adding 2+2 dimensions is a minimal prescription which is sufficient to deal with bosons, but 4+4 is necessary to deal with fermions [8], and is therefore useful for supersymmetry. The extra fields introduced by adding 4+4 instead of 2+2 are thus nonminimal, and our treatment of the NSR superstring will be analogous.

For example, for Yang-Mills this means starting with a $(D-2)$-component vector $|A\rangle = A_r|r\rangle$ ($r = 1, ..., D-2$) and extending the basis $|r\rangle$ ($\langle r|s\rangle = \delta^{rs}$) so that $r$ runs over $(D-2)+2n$ commuting values and $2n$ anticommuting. The transverse Lorentz spin operators $M^{rs} = ||r\rangle\langle s|| \equiv |r\rangle\langle s| - |s\rangle\langle r|$ are then generalized in the same way. That means that $M^{rs}$ gets generalized from SO(D-2) spin to OSp(D,2|4) (the graded orthogonal group for D space, 2 time, and 4 fermionic dimensions). (This OSp group, as well as the other OSp groups discussed in this section, applies to both particles and strings, and should not be confused with the OSp(1|2) subgroup of 2D superconformal transformations discussed elsewhere in this paper, which applies only to strings.) We divide up this $(D,2|4)$ dimensional basis into $(D-1,1|2)+(1,1|2)$ as $|i\rangle = (|a\rangle, |\alpha\rangle)$ and $|A\rangle = (|\pm\rangle, |\alpha'\rangle)$, and write the anticommuting values of the Sp(2) index $\alpha$ as $\alpha = (c, \tilde{c})$. "$a$" is the usual physical SO(D-1,1) index, and $\pm$ is an SO(1,1) index. We then have the generators of the corresponding spin group OSp(D−1,1|2)⊗OSp(1,1|2): $M^{ij} = ||i\rangle\langle j||$ and $S^{AB} = ||A\rangle\langle B||$. $M^{ij}$ are the spin operators of the "minimal" case, adding 2+2 dimensions to the light cone, corresponding to the usual quantization of the NSR superstring, while $S^{AB}$ are the "nonminimal" spin generators, corresponding to the additional degrees of freedom we will add for the NSR superstring.

The BRST and ghost number operators then take the form, for arbitrary massless particles (generalization to the massive case is easy),

$$Q = c\tfrac{1}{2}p^2 + i(M^{ca}p_a + S_-{}^{c'}) - M^{cc}b \quad, \quad J = \tfrac{1}{2}[c,b] + i(M^{c\tilde{c}} + S^{c'\tilde{c}'}) \, . \tag{3.1}$$

In the case of Yang-Mills, we can then straightforwardly derive gauge transformations $\delta|A\rangle = Q|\lambda\rangle$ and the action $\tfrac{1}{2}\langle A|QA\rangle$ for the free theory, with the expansion of the field $|A\rangle$ and some useful equations:

$$|A\rangle = A_a|a\rangle + iC|0\rangle - i\tilde{C}|c\rangle$$
$$+ iC'|\tilde{c}'\rangle - i\tilde{C}'|c'\rangle + A_+|+\rangle + A_-|-\rangle$$
$$+ iB_a c|a\rangle - Bc|0\rangle + \tilde{B}c|c\rangle$$
$$- B'c|\tilde{c}'\rangle + \tilde{B}'c|c'\rangle + iB_+c|+\rangle + iB_-c|-\rangle \quad,$$

$$\langle a|b\rangle = \eta^{ab} \quad, \quad \langle +|-\rangle = 1 \quad, \quad \langle c|0\rangle = \langle c'|\tilde{c}'\rangle = i \quad,$$
$$J = (-1, 1, -1, 1) \quad \text{on} \quad (|0\rangle, |c\rangle, |\tilde{c}'\rangle, |c'\rangle) \quad,$$
$$M^{cc}|0\rangle = 2i|c\rangle \quad, \quad M^{ca}|0\rangle = -i|a\rangle \quad, \quad M^{ca}|b\rangle = \eta^{ab}|c\rangle \quad,$$
$$S_-{}^{c'}|\tilde{c}'\rangle = i|+\rangle \quad, \quad S_-{}^{c'}|-\rangle = -|c'\rangle \quad, \tag{3.2}$$

where the spin operators appearing in (3.1) vanish on other states. We have explicitly identified the state $|\tilde{c}\rangle$ for the Yang-Mills ghost as the "vacuum" $|0\rangle$. Details of the interacting theory (without the nonminimal fields $C'$, $\tilde{C}'$, and $A_\pm$) can be found in [7]. We will return below to interactions by the method of external fields in order to stress the relation of this formulation of Yang-Mills to the usual formulations of string theory.

### 3.2. First quantization of Dirac spinor

For the Dirac spinor (and similarly for NSR), there are three simple ways to derive the first-quantized BRST operator: (1) Add extra dimensions to the light cone. This method works for both particles and strings. (The other two methods work for strings and a few kinds of particles, but not for Yang-Mills.) For this case, we add 2+2 dimensions for the $x$ coordinates and 4+4 for the $\psi$'s. (2) Start with the constraint algebra + an extra abelian constraint, and write the BRST operator for these constraints. (3) First-quantize the classical mechanics action. We will describe this last method, since it most closely relates to the super-Riemann surface approach, and may give some insight into it. The other two approaches are actually simpler, since the nonminimal term can be treated separately. In all cases, the result for $Q$ will be simply to add a quadratic nonminimal term to the usual minimal $Q$.

The easiest way to derive lagrangians for the mechanics of a relativistic system is to start with the hamiltonian form of the lagrangian, which simply states which variables are canonically conjugate and what the constraints are: For the spinor,

$$L_H = (\dot{x} \cdot p + i\tfrac{1}{2}\dot{\psi} \cdot \psi) - (g\tfrac{1}{2}p^2 + i\chi\psi \cdot p) \quad . \tag{3.3}$$

The gauge transformations of all the variables except the Lagrange multipliers are given by commutators with the gauge generators (constraints), while the Lagrange multipliers transform exactly like the time components of gauge fields (as derived e.g., by covariantizing the time derivative by adding to it the latter term in $L_H$). (For a review and examples, see e.g., [9].)

Gauge fixing for reparametrizations works the same way as for the spinless particle; here we focus mainly on the 1D supersymmetry. The lagrangian BRST transformations on the physical fields $x$, $p$, and $\psi$, the gauge fields $g$ and $\chi$, the ghosts $c$ and $\gamma$, antighosts $b$ and $\beta$, and NL fields $B$ and $\mu$ are

$$Qx = icp + i\gamma\psi \quad, \quad Q\psi = \gamma p \quad, \quad Qp = 0 \quad,$$
$$Qg = \dot{c} + 2\gamma\chi \quad, \quad Qc = -\gamma^2 \quad,$$
$$Qb = B \quad, \quad QB = 0 \quad,$$
$$Q\chi = \dot{\gamma} \quad, \quad Q\gamma = 0 \quad,$$
$$Q\beta = -\mu \quad, \quad Q\mu = 0 \quad . \tag{3.4}$$

The temporal gauge $\chi = 0$ is then chosen by

$$\Psi = \int ib(g-1) + i\beta\chi$$

$$\Rightarrow \quad \Delta L = Q\Psi = \int iB(g-1) - ib(\dot{c} + 2\gamma\chi) - i\mu\chi + i\beta\dot{\gamma}$$

$$\Rightarrow \quad L_{gf} = (\dot{x} \cdot p + i\tfrac{1}{2}\dot{\psi} \cdot \psi + i\dot{c}b + i\dot{\gamma}\beta) - \tfrac{1}{2}p^2 \quad , \qquad (3.5)$$

where we have eliminated auxiliary fields by their equations of motion in the final result.

To fix Lorentz gauges in a first-order formalism, we introduce a quartet of nonminimal fields (the defining representation of OSp(1,1|2), which includes BRST as a subgroup) by gauge fixing a trivial gauge invariance: We introduce a field $\tilde{\chi}$ which is pure gauge, and thus does not appear in the gauge invariant lagrangian, along with the corresponding ghost $\tilde{\gamma}$, antighost $\tilde{\beta}$, and NL field $\tilde{\mu}$:

$$Q\tilde{\chi} = \tilde{\gamma} \quad , \quad Q\tilde{\gamma} = 0 \quad ,$$
$$Q\tilde{\beta} = -\tilde{\mu} \quad , \quad Q\tilde{\mu} = 0 \quad . \qquad (3.6)$$

We then consider the following gauges:

$$\text{temporal:} \quad \Psi_1 = \int ib(g-1) + i\beta\chi + i\tilde{\beta}\tilde{\chi}$$

$$\text{"simple" Lorentz:} \quad \Psi_2 = \int ib(g-1) + i\beta\chi + i\tilde{\beta}\dot{\tilde{\chi}}$$

$$\text{"normal" Lorentz:} \quad \Psi_3 = \int ib(g-1) + i\beta(\chi - \tilde{\chi}) + i\tilde{\beta}\dot{\tilde{\chi}} \quad ,$$
$$(3.7)$$

which lead to the gauge fixing terms

$$Q\Psi_1 = \int iB(g-1) - ib(\dot{c} + 2\gamma\chi) - i\mu\chi + i\beta\dot{\gamma} - i\tilde{\mu}\tilde{\chi} + i\tilde{\beta}\tilde{\gamma}$$

$$Q\Psi_2 = \int iB(g-1) - ib(\dot{c} + 2\gamma\chi) - i\mu\chi + i\beta\dot{\gamma} - i\tilde{\mu}\dot{\tilde{\chi}} + i\tilde{\beta}\dot{\tilde{\gamma}}$$

$$Q\Psi_3 = \int iB(g-1) - ib(\dot{c} + 2\gamma\chi) - i\mu(\tilde{\chi} - \chi) + i\beta(\dot{\gamma} - \tilde{\gamma})$$
$$- i\tilde{\mu}\dot{\tilde{\chi}} + i\tilde{\beta}\dot{\tilde{\gamma}} \quad , \qquad (3.8)$$

and the gauge fixed lagrangians

$$L_1 = (\dot{x} \cdot p + i\tfrac{1}{2}\dot{\psi} \cdot \psi + i\dot{c}b + i\dot{\gamma}\beta) - \tfrac{1}{2}p^2$$
$$L_2 = (\dot{x} \cdot p + i\tfrac{1}{2}\dot{\psi} \cdot \psi + i\dot{c}b + i\dot{\gamma}\beta + i\dot{\tilde{\gamma}}\tilde{\beta} + i\dot{\tilde{\chi}}\tilde{\mu}) - \tfrac{1}{2}p^2$$
$$L_3 = (\dot{x} \cdot p + i\tfrac{1}{2}\dot{\psi} \cdot \psi + i\dot{c}b + i\dot{\gamma}\beta + i\dot{\tilde{\gamma}}\tilde{\beta} + i\dot{\chi}\tilde{\mu})$$
$$- [(\tfrac{1}{2}p^2 + i\tilde{\gamma}\beta) + i\chi(\psi \cdot p - 2\gamma b)] \quad . \qquad (3.9)$$

The temporal gauge is $\chi = \tilde{\chi} = 0$, the simple Lorentz gauge is $\chi = 0$, $\dot{\tilde{\chi}} = 0$, and the normal Lorentz gauge is $\chi = \tilde{\chi}$, $\dot{\tilde{\chi}} = 0$. The normal Lorentz gauge thus sets $\dot{\chi} = 0$, so the gravitino is propagating.

### 3.3. BRST for Dirac spinor

The resulting hamiltonian BRST operators are

$$Q_{min} = Q_1 = c\tfrac{1}{2}p^2 + \gamma\psi \cdot p - \gamma^2 b \quad ,$$
$$Q_{nonmin} = Q_2 = Q_3 = Q_1 + \tilde{\gamma}\tilde{\mu} \quad . \qquad (3.10)$$

These are the usual BRST operators following from the hamiltonian treatment of the constraints, where the Lorentz gauges differ by just a nonminimal term. The gauge fixed lagrangians differ only by: (1) minimal vs. nonminimal fields in the term defining canonical conjugates, and (2) a different choice of hamiltonian gauge fermion for the two Lorentz gauges: In the language of (2.7), $f_1 = f_2 = 0$, $f_3 = i\beta\chi$. (From now on, we drop the $\tilde{\phantom{x}}$'s on $\chi$ and $\mu$.)

As discussed in [1], the assignment to the wave functions of proper boundary conditions in the ghost coordinates allows the use of $\Phi^\dagger Q\Phi$ actions in the second-quantized theory, without the use of picture changing operators. In this case, the Hilbert space is defined with respect to the bosonic ghosts by interpreting $\frac{1}{\sqrt{2}}(\gamma + i\tilde{\gamma})$ and $\frac{1}{\sqrt{2}}(\beta + i\tilde{\beta})$ as creation operators and $\frac{1}{\sqrt{2}}(\gamma - i\tilde{\gamma})$ and $\frac{1}{\sqrt{2}}(\beta - i\tilde{\beta})$ as annihilation operators. This has a natural interpretation in terms of OSp(D,2|4) [8]: $(\psi_a, \chi, \mu; \gamma, \beta, \tilde{\gamma}, \tilde{\beta})$ are Dirac matrices for OSp(D,2|4), with a corresponding infinite-dimensional spinor representation. We then recognize $Q_{nonmin}$ as being of the form (3.1), where the OSp(D,2|4) spin operators are quadratic in these OSp(D,2|4) Dirac matrices. The usual physical components are those which are singlets under the Sp(2) subgroup generated by $M^{\alpha\beta} + S^{\alpha'\beta'}$. The minimal ghosts, corresponding to adding just 2+2 dimensions to get OSp(D-1,1|2), give a spinor which does not contain such singlets.

To analyze the cohomology of $Q$, we first make a field redefinition $\Phi' = exp(iS_+^{c'}b)\Phi$, which is equivalent to the transformation

$$Q' = e^{-iS_+^{c'}b}Q e^{iS_+^{c'}b}$$
$$= c\tfrac{1}{2}p^2 + i(M^{ca}p_a + S_-^{c'} + S_+^{c'}\tfrac{1}{2}p^2) - (M^{cc} + S^{c'c'})b \quad , (3.11)$$

where in this case $S_+^{c'} = i\tilde{\gamma}\chi$. This is the form in which $Q$ directly appears as one of the generators of OSp(1,1|2) with nonminimal fields, rather than the form (3.1), which is simpler when one ignores antiBRST [8]. (OSp(1,1|2) is an extension of BRST to include antiBRST in a way which is useful for second-quantization via first-quantized BRST.) Eliminating the antifields (those not annihilated by $b$) by the equations of motion following from (3.11) generally leaves just the BRST-Sp(2) singlets: in this case, the SO(D,2) spinor representing $(\psi_a, \chi, \mu)$. (Similar remarks apply for the vector.) The nonminimal term in the BRST operator then allows the gauge choice $\chi = 0$. The remaining spinor field $|s\rangle$, which is a representation of $\psi$ and $x$ only, satisfies

$$\frac{1}{\sqrt{2}}(\gamma - i\tilde{\gamma})|s\rangle = \frac{1}{\sqrt{2}}(\beta - i\tilde{\beta})|s\rangle = \chi|s\rangle = b|s\rangle = 0 \quad . \qquad (3.12)$$

The antifield ($c$) part of the solution to the cohomology can then easily be found in terms of this by re-solving $Q = 0$. Finally, inverting the transformation in (3.11), we find the solution to the cohomology for the original $Q$:

$$|\Phi\rangle = (1 - \gamma\beta + ic\mu\beta)|s\rangle$$
$$= \left[\tfrac{3}{2} - \tfrac{1}{4}(\gamma + i\tilde{\gamma})(\beta + i\tilde{\beta}) + \tfrac{1}{2}ic\mu(\beta + i\tilde{\beta})\right]|s\rangle \quad ,$$
$$p^2|s\rangle = \psi \cdot p|s\rangle = 0 \quad . \quad (3.13)$$

As usual, $|s\rangle$ can be expressed as a plane wave $u^\alpha(k)e^{-ik\cdot x}|\alpha\rangle$ in terms of the solution $u$ to the Dirac equation and a basis $|\alpha\rangle$ for representing $\psi$.

If we choose a coordinate basis for the wave functions (fields), the more convenient choices are ones in which the wave functions can be chosen to be real. In such a representation the creation and annihilation operators must be real (up to an overall phase), as for the usual harmonic oscillator ($a = x + \partial/\partial x \rightarrow |0\rangle \sim e^{-x^2/2}$). In this case the real wave functions are $\Phi(\gamma, \tilde{\beta})$, so $|s\rangle \sim e^{\gamma\tilde{\beta}}$ (or similarly for $\Phi(\beta, \tilde{\gamma})$).

### 3.4. Second quantization

In general, $\Phi^\dagger Q\Phi$ actions have two particularly useful gauges: (1) a "Wess-Zumino"-type gauge, where only the usual physical fields remain, which is useful for comparison with standard formulations of the field theory, and (2) a "Fermi-Feynman"-type gauge, where propagators are proportional to $1/(p^2 + M^2)$, which is useful for perturbation theory. This is the same situation that occurs in the superfield formulation of 4D N=1 supersymmetric particle field theories. In fact, some of the additional fields appearing at the massless level in the NSR superstring also appear in the superfield formulation of 4D N=1 super Yang-Mills [8]: If we eliminate the ghosts (leaving just the physical, gauge, and auxiliary fields which appear in the gauge invariant lagrangian) by looking at just the SO(D,2) subgroup of OSp(D,2|4), then the SO(D,2) vector and spinor representing super Yang-Mills break up into SO(D-1,1)⊗SO(1,1) representations as:

| SO(1,1) charge | SO(D-1,1) field |
|---|---|
| $+1$ : | $A_-$ (auxiliary scalar) |
| $+\tfrac{1}{2}$ : | $\lambda^\alpha$ (physical spinor) |
| $0$ : | $A_a$ (physical vector) |
| $-\tfrac{1}{2}$ : | $\kappa_\alpha$ (gauge spinor) |
| $-1$ : | $A_+$ (gauge scalar) |

The engineering dimension of the fields is also given by the SO(1,1) charge up to a constant.

To gauge fix the $\Phi^\dagger Q\Phi$ action for the spinor to a WZ gauge it is sufficient to look at just the SO(D,2) spinor, consisting of $\lambda$ and $\kappa$ above, since these are all that remain after eliminating the antifields by their nondynamical equations of motion. In terms of the anticommuting coordinate $\mu$, these can be represented as $|\Phi(\mu)\rangle = |\lambda\rangle + \mu|\kappa\rangle$, where now $\frac{1}{\sqrt{2}}(\gamma - i\tilde{\gamma})\Phi = \frac{1}{\sqrt{2}}(\beta - i\tilde{\beta})\Phi = 0$. As can be seen by the gauge invariance $\delta\Phi = Q\Lambda$, these fields can only appear in the gauge invariant combination $|\lambda'\rangle = |\lambda\rangle + i\psi \cdot p|\kappa\rangle$. In terms of this gauge invariant field (or equivalently choosing the gauge $\chi|\Phi\rangle = 0$) the lagrangian reduces to just the usual $\bar{\lambda}'\psi \cdot p\lambda'$.

Gauge fixing to the FF gauge is much simpler: According to (2.8), just set $b|\Phi\rangle = 0$. This gauge is just the covariantized light cone one, which results from adding 4+4 dimensions to the light-cone Feynman rules, and gives results equivalent to the light-cone ones by the usual Parisi-Sourlas arguments (at least for the case of doing one-loop calculations by putting this spinor in an external field).

### 3.5. External fields

These BRST operators can easily be generalized to include external fields: For example, for external Yang-Mills, the Dirac equation constraint covariantizes to $\psi \cdot (p + A)$, and the Klein-Gordon equation constraint is just the square of this. The result is, for the minimal case,

$$Q_{int} = c[\tfrac{1}{2}(p + A)^2 + \tfrac{1}{2}\sigma^{ab}F_{ab}] + \gamma\psi \cdot (p + A) - \gamma^2 b \quad , \quad (3.14)$$

where $\sigma^{ab} = \tfrac{1}{2}[\psi^a, \psi^b]$ is the spin and $F_{ab} = [(p+A)_a, (p+A)_b]$ is the Yang-Mills field strength. If we expand $Q_{int}$ in $A$ as $Q_{int} = Q + V + \cdots$ and choose $A_a(x)$ to be a momentum eigenstate $\epsilon_a e^{-ik\cdot x}$ ($k \cdot \epsilon = 0$), we find the vertex operator

$$V = \epsilon^a(\gamma\psi + cp + ck\cdot\psi\psi)_a e^{-ik\cdot x} \quad . \quad (3.15)$$

For Yang-Mills in external Yang-Mills, BRST covariantizes to

$$Q_{int} = c[\tfrac{1}{2}(p+A)^2 + \tfrac{1}{2}M^{ab}F_{ab}] + iM^{ca}(p+A)_a - M^{cc}b \quad . \quad (3.16)$$

(This expression also gives (3.14) if we substitute the OSp spin operators for the spinor instead of the vector, as expected from supersymmetry.) The vertex operator is now

$$V = \epsilon^a(iM^c{}_a + cp_a + ck^b M_{ba})e^{-ik\cdot x} \quad . \quad (3.17)$$

(Again, this gives (3.15) with the appropriate substitutions.) However, in the Yang-Mills case, there is an interesting analogy to strings because of the occurence of the "vacuum" $|0\rangle$ in the spectrum: We can write the physical Yang-Mills state in terms of the vertex operator acting on it,

$$V|0\rangle = \epsilon_a e^{-ik\cdot x}|a\rangle \quad , \quad (3.18)$$

and the 3-string vertex can be written as

$$\langle 0|VVV|0\rangle \quad . \tag{3.19}$$

Higher-point functions can also be treated, though not as conveniently as in string theory, where duality allows all the diagrams to be expressed as a single propagator with insertions of 3-point vertices. Using the Fermi-Feynman gauge propagator $b/p^2$, a diagram can be written as (using $b|0\rangle = 0$)

$$\langle 0|VV\frac{b}{p^2}V\cdots\frac{b}{p^2}VV|0\rangle = \langle 0|VV\frac{1}{p^2}\mathcal{W}\cdots\frac{1}{p^2}\mathcal{W}\{\mathcal{W},V\}|0\rangle \quad ,$$
$$\mathcal{W} = \{b,V\} \quad , \tag{3.20a}$$

where in the present case of Yang-Mills, to lowest order in external fields in each vertex,

$$\mathcal{W} = \epsilon^a(p_a + k^b M_{ba})e^{-ik\cdot x} \tag{3.20b}$$

consists of the usual orbital + spin interaction. If we eliminate the remaining ghost degrees of freedom in (3.20a), the result agrees with that which follows from simply doing a perturbation expansion of $H_{int}$ of (2.12).

We can also easily check relations such as (2.11). Furthermore, unlike the spin-0 and $\frac{1}{2}$ cases of $M^{ij}$ for (3.16), for self-interacting Yang-Mills $Q_{int}^2 = 0$ only when the external field is on shell.

## 4. FIRST QUANTIZATION OF NSR

As for the particle, we start with the hamiltonian form of the lagrangian. In the case of the NSR (nonheterotic) string, knowing the commutation relations and constraints, we can directly write

$$L_H = (\dot{x}\cdot\pi + i\tfrac{1}{2}\dot{\psi}_\pm\cdot\psi_\pm) - \left[g_\mp(\tfrac{1}{2}p_\pm^2 \pm i\tfrac{1}{2}\psi'_\pm\cdot\psi_\pm) + i\chi_\mp\psi_\pm\cdot p_\pm\right] \quad , \tag{4.1}$$

where $p_\pm \equiv (\pi \pm x')/\sqrt{2}$ and $' \equiv \partial/\partial\sigma$. This is the simplest form from which to derive hamiltonian BRST operators: It avoids messy field redefinitions needed to simplify the more nonlinear expressions obtained from quantizing the usual second-order lagrangians in the usual Lorentz gauges (see [10] for the case of the Veneziano string). It is also the simplest way to derive the usual second-order lagrangian: Simply eliminate the auxiliary field $\pi$ by its equation of motion (in both the lagrangian and transformation laws). We then find the usual result, including quartic fermion terms:

$$L = -\,\mathrm{e}^{-1}(e_+x)\cdot(e_-x) \mp i\tfrac{1}{2}\psi_\pm e_\mp \psi_\pm$$
$$\pm i\mathrm{e}^{-1}\chi_\pm\psi_\mp\cdot e_\mp x - \tfrac{1}{2}\mathrm{e}^{-1}\chi_+\chi_-\psi_+\psi_- \quad , \tag{4.2}$$

where we use (one-component) Weyl spinor notation, all derivatives have been collected into the zweibein (in a particular Weyl scale and local Lorentz gauge) as

$$e_\pm \equiv e_\pm{}^m \partial_m = g_\pm \partial_1 \pm \partial_0 \quad , \quad \mathrm{e} \equiv det\, e_a{}^m = g_+ + g_- \quad , \tag{4.3}$$

and the fermions have been rescaled by convenient powers of e.

Gauge transformations can be derived by the same methods as for the particle. Here we will instead write down the BRST operator directly from the constraints; the relation to the lagrangian approach follows the analysis of section 3.2. As there for the particle, the addition of a nonminimal term gives the same result, up to a unitary transformation, as gauge fixing the gravitino in a Lorentz gauge, so that one of the nonminimal fields can be interpreted as a propagating gravitino. The result is

$$Q = \int c(\tfrac{1}{2}p^2 + \tfrac{1}{2}\psi'\cdot\psi + c'b + \tfrac{3}{2}\gamma'\beta + \tfrac{1}{2}\gamma\beta') + \gamma\psi\cdot p - \gamma^2 b + \tilde{\gamma}\mu \quad , \tag{4.4}$$

where now $\int \equiv \oint dz/2\pi i$ and $' \equiv d/dz$. (As usual, we have defined $\rho = \tau + i\sigma$ and conformally transformed from $\rho$ to $z$, where $z = e^\rho$ near the end of a string, or on the free string.) The operators are normalized so that the singular term in the operator product $b(z_1)c(z_2)$ is $1/(z_1 - z_2)$, and the same for $\psi\psi$, $\chi\mu$, $\beta\gamma$, and $\tilde{\beta}\tilde{\gamma}$, while for $pp$ it is $1/(z_1-z_2)^2$. The Virasoro operators can be defined by the generalization of (2.6,7): If we define them as $\{Q,b\}$ as in (2.6), the nonminimal fields will be invariant under the corresponding conformal transformations. We therefore choose, as in (2.7), the Virasoro operators

$$\begin{aligned}\mathbf{L} &\equiv \{Q, b+f\} = \{Q, b + \tfrac{1}{2}\tilde{\beta}\overset{\leftrightarrow}{\partial}\chi + w(\tilde{\beta}\chi)'\} \\ &= \tfrac{1}{2}p^2 + \tfrac{1}{2}\psi'\cdot\psi + 2c'b + cb' + \tfrac{3}{2}\gamma'\beta + \tfrac{1}{2}\gamma\beta' \\ &\quad + (w+\tfrac{1}{2})\tilde{\gamma}'\tilde{\beta} + (w-\tfrac{1}{2})\tilde{\gamma}\tilde{\beta}' + (w+\tfrac{1}{2})\chi'\mu + (w-\tfrac{1}{2})\chi\mu' \quad ,\end{aligned} \tag{4.5}$$

where we have chosen $f$ so that the nonminimal fields are conformal with weights $\tfrac{1}{2} \pm w$. $\mathbf{L}_0$ is still of the form $p^2 + M^2$, but now $M^2$ depends on the mode numbers of the nonminimal coordinates as well as the minimal ones. As in section 3 for the gauge-fixed lagrangian, this corresponds to modifying the gauge fixing condition: $[e_\pm - (w-\tfrac{1}{2})\omega_\pm]\chi_\pm = 0$, where $\omega_\pm \equiv \partial_m e_\pm{}^m$ is the Lorentz connection (actually $\pm\omega_\pm$). Although the conformal weights $\tfrac{1}{2} \pm w$ for the nonminimal fields are arbitrary, we choose $w=1$ so that $\gamma \pm i\tilde{\gamma}$ and $\beta \pm i\tilde{\beta}$ have definite conformal weight.

This procedure can be applied in arbitrary theories to give arbitrary conformal weights to nonminimal fields. In particular, the conformal weights assigned in [11] to the nonminimal fields in the Green-Schwarz formulation of the superstring are consistent by such a procedure with those later chosen in [12], which explains why they both gave a vanishing conformal anomaly. (The former choice has the advantage that the Virasoro operators commute with the BRST-Sp(2) that rotates ghosts into antighosts.)

The boundary conditions in the Neveu-Schwarz sector with respect to the ghosts follow from just treating the positive-energy oscillators as creation operators and the negative-energy ones as annihilation operators. For the bosonic oscillators, this prevents the occurence of states of arbitrarily negative energy. The same holds for the Ramond sector, except that there are now also zero-energy modes, which are just the modes of the spinor particle described in section 3, and are therefore treated the same way. For example, in the NS sector, the state of lowest energy (i.e., (mass)$^2$) is the tachyon, which satisfies

$$b_0|t\rangle = 0 \quad ,$$

$$\zeta_n|t\rangle = 0 \quad \forall n > 0 \quad \forall \text{ oscillators } \zeta \qquad (4.6a)$$

The Yang-Mills ghost, which we'll treat as the first-quantized "vacuum," is given by

$$|0\rangle = \beta_{-1/2}|t\rangle \quad , \qquad (4.6b)$$

and the physical Yang-Mills field as usual by

$$|a\rangle = \psi^a_{-1/2}|t\rangle \quad . \qquad (4.6c)$$

In the R sector, the massless spinor satisfies $\zeta_n|s\rangle = 0$ for all $n > 0$ for all oscillators $\zeta$, while the $\zeta_0$ act as described in section 3.3.

## 5. CONFORMAL FIELD THEORY

### 5.1. NS Vertex operators

The particle Yang-Mills vertex was obtained in section 3.5 by covariantizing the derivative $p$, and this method can also be applied to strings. Another way is to consider $[Q, x_a]$, which will pick out the same terms at $k = 0$, multiply by $e^{-ik\cdot x}$, and add corrections perturbatively in $k$ until a BRST invariant expression is obtained. An equivalent way is to consider $[Q, p_a]$, and first pull off an overall derivative before following the same procedure. This is characteristic of a general relation between two types of vertex operators in string theory: vertex operators which are integrated and have ghost number zero, and those which are local on the world sheet and have ghost number one, which are more useful in the second-quantized approach. The relation is:

$$[Q, \mathcal{W}] = 0 \quad , \quad \mathcal{W} = \int W$$

$$\Rightarrow \quad [Q, W] = V' \quad \Rightarrow \quad \{Q, V\} = 0 \quad . \qquad (5.1)$$

The last equation follows from the fact that $\{Q, V\}$ must have vanishing derivative but is local by construction, and therefore must vanish. $\mathcal{W}$ is then the integrated form of the vertex, while $V$ is the local version. This $\mathcal{W}$ is the same one which appeared in (2.12): It is just the contribution of the interactions to $H_{int} = \mathbf{L}_0 + \mathcal{W}$. The ambiguity $\delta W = \lambda'$, which has the form of a (1D) gauge transformation, is equivalent to the BRST-type gauge transformation $\delta V = [Q, \lambda]$. Without loss of generality, $W$ can be chosen to have conformal weight 1 with respect to $\mathbf{L} \equiv \{Q, b + f\}$, so $\mathcal{W}$ is conformally invariant. Then $V$ has conformal weight 0. We can then invert the derivation (5.1) of $V$ from $W$ as (see (2.12))

$$W = \left\{ \int (b+f), V \right\}$$

$$\Rightarrow \quad [Q, W] = \left[ \int \mathbf{L}, V \right] = V' \quad . \qquad (5.2)$$

This works for general conformal coordinates (with the $\int$ and $'$ for that coordinate): In particular, for the $\sigma$ coordinate $\int \mathbf{L} = \mathbf{L}_0$ (but in the $z$-plane $\int \mathbf{L} = \mathbf{L}_{-1}$), etc. Remember that $W$ depends on the gauge choice, fixed by $f$. This appears more subtly in (5.1): Since $\mathbf{L}$ also depends on $f$, the notion of "locality" in the expression $\mathcal{W} = \int W$ depends on how conformal transformations are defined. The class of $f$'s defined by (4.5) implies that $W$ be local in the nonminimal coordinates as well as the minimal ones to transform simply under the corresponding conformal generators $\mathbf{L}$.

In the Veneziano string theory $W$ is some operator constructed from $x$, and $V$ is just $cW$ (e.g., $W = pe^{-ik\cdot x}$ for the vector). This can be generalized easily to the NS string. We can again use (5.1), or we can go to world-sheet superspace:

$$[Q, \mathcal{W}] = 0 \quad , \quad \mathcal{W} = \oint \frac{dz}{2\pi i} d\theta \, \mathbf{W}$$

$$\Rightarrow \quad \{Q, \mathbf{W}\} = -\mathbf{DV} \quad \Rightarrow \quad \{Q, \mathbf{V}\} = 0$$

$$\Rightarrow \quad [Q, W] = \int d\theta \, \mathbf{DV} = V' \quad \Rightarrow \quad V = \mathbf{V}|_{\theta=0} \quad , \qquad (5.3)$$

where $\mathbf{D} = \partial/\partial\theta + \theta\partial/\partial z$. This is just a reflection of the fact that $\mathbf{V}$ differs at different points in superspace only by a BRST variation, the analog of the statement that $\mathbf{W}$ can be chosen superconformal covariant with weight $\frac{1}{2}$ (so $\mathcal{W}$ is a superconformal invariant).

Thus physical NS vertex operators can always be expressed as

$$V = -\int d\theta \, \mathbf{CW}[k, \mathbf{X}(\mathbf{Z})] \quad \text{for some } \mathbf{W} \quad , \quad \mathbf{Z} = (z, \theta) \quad ,$$

$$\mathbf{C} = c - \theta\gamma, \quad \mathbf{X} = \hat{x} + \theta\psi, \quad \mathbf{DX} = \psi + \theta p \quad ,$$

$$p = \hat{x}' \quad , \quad x(z, \bar{z}) = i[\hat{x}(z) - \hat{\bar{x}}(\bar{z})] \quad . \qquad (5.4)$$

We have here performed a minor slight-of-hand: By the above analysis, we should actually get $\mathbf{CDW} - \frac{1}{2}(\mathbf{DC})\mathbf{W}$ for $\mathbf{V}$, but this evaluated at $\theta = 0$ (for $V$) becomes just $-\int d\theta \, \mathbf{CW}$ by redefining $\mathbf{C}$ to have an extra $\frac{1}{2}$ in its $\theta$ term.

As an example, we have the Yang-Mills vertex [13] (cf. (3.15)):

$$\mathbf{W}(k, \mathbf{Z}) = \epsilon^a(\mathbf{DX}_a)e^{k\cdot\mathbf{X}} \quad \Rightarrow$$

$$\mathcal{W} = \oint \frac{dz}{2\pi i} \, \epsilon^a(p + k\cdot\psi\psi)_a e^{k\cdot\hat{x}} \quad ,$$

$$V = -\epsilon^a \int d\theta \, (\mathbf{CDX})_a e^{k\cdot\mathbf{X}} = \epsilon^a (\gamma\psi + cp + ck\cdot\psi\psi)_a e^{k\cdot\hat{x}} \quad . \quad (5.5)$$

## 5.2. Bosonization and OSp(1|2)

Because of our choice of boundary conditions, our bosonization of the bosonic ghosts differs from the usual [13]:

$$\frac{1}{\sqrt{2}}(\gamma - i\tilde{\gamma}) = \eta e^\phi \quad , \quad \frac{1}{\sqrt{2}}(\beta + i\tilde{\beta}) = -\xi' e^{-\phi} \quad ,$$

$$\frac{1}{\sqrt{2}}(\gamma + i\tilde{\gamma}) = \bar{\eta} e^{\bar{\phi}} \quad , \quad \frac{1}{\sqrt{2}}(\beta - i\tilde{\beta}) = -\bar{\xi}' e^{-\bar{\phi}} \quad ,$$

$$\chi = e^{-\widetilde{\phi}} \quad , \quad \mu = e^{\widetilde{\phi}} \quad . \quad (5.6)$$

(There are also some Klein transformations/Jordan-Wigner factors/cocycles, which we have carelessly omitted.) As usual, $e^{\alpha\cdot\zeta}$ has conformal weight $\frac{1}{2}\alpha\cdot\alpha + w\cdot\alpha$, and operator products satisfy $e^{\alpha_1\cdot\zeta(z_1)}e^{\alpha_2\cdot\zeta(z_2)} = e^{\alpha_1\cdot\zeta(z_1)+\alpha_2\cdot\zeta(z_2)}(z_1-z_2)^{\alpha_1\cdot\alpha_2}$, where $\alpha\cdot\zeta \equiv \alpha^i\zeta_i$ and $\frac{1}{2}\alpha\cdot\alpha + w\cdot\alpha \equiv \eta_{ij}(\frac{1}{2}\alpha^i\alpha^j + w^i\alpha^j)$, $\eta_{ii} = +1$ for bosonized fermions and $\hat{x}$ and $-1$ for bosonized bosons, and $w^i = 0$ for physical spinors and $\hat{x}$ and 1 for these ghosts. (It's $\frac{3}{2}$ for $c$ and $b$, but we don't bosonize those here.)

Another way to look at the NS vertex construction is in terms of the OSp(1|2)-invariant vacuum $|OSp\rangle$: Although it is not in the Hilbert space, we can formally write states in terms of it, then "picture-change" the vertex operators back in terms of the Yang-Mills ghost vacuum $|0\rangle$. In the original NSR formalism without nonminimal coordinates, the relation between these two vacuua was simply $|OSp\rangle = X|0\rangle$ in terms of the picture-changing operator $X$, or $|0\rangle = Y|OSp\rangle$ in terms of the inverse picture-changing operator. One way to define $X$ is as $\{Q, \xi\}$ (not a true BRST commutator, since $\xi_0$ is not really defined in terms of $\beta$ and $\gamma$). However, in (5.6) we now have both a $\xi$ and a $\bar{\xi}$, and therefore the corresponding modification to this picture changing is:

$$X = \{Q, \xi\} \equiv :\frac{1}{\sqrt{2}} e^\phi ([\beta, Q] + i\mu):$$
$$= e^\phi \frac{1}{\sqrt{2}}(\psi\cdot p + i\mu) + \tfrac{1}{2} c\xi' + \tfrac{1}{2}\eta' e^{2\phi} b + \tfrac{1}{2}(\eta e^{2\phi} b)'$$
$$\quad + \tfrac{3}{4} c'\bar{\xi}' e^{\phi-\bar{\phi}} + \tfrac{1}{2} c(\bar{\xi}' e^{-\bar{\phi}})' e^\phi - \bar{\eta} e^{\phi+\bar{\phi}} b \quad ,$$
$$\overline{X} = \{Q, \bar{\xi}\} \equiv :\frac{1}{\sqrt{2}} e^{\bar{\phi}}([\beta, Q] - i\mu): \quad ,$$
$$Y = \frac{\sqrt{2}}{4}(3 e^{-\phi-\widetilde{\phi}} - i\sqrt{2} c\xi' e^{-2\phi} - \bar{\eta}\xi' e^{-2\phi+\bar{\phi}-\widetilde{\phi}} + \tfrac{1}{2} cc'\xi'\xi'' e^{-3\phi-\widetilde{\phi}}) \quad ,$$
$$\overline{Y} = \frac{\sqrt{2}}{4}(3 e^{-\bar{\phi}-\widetilde{\phi}} + i\sqrt{2} c\bar{\xi}' e^{-2\bar{\phi}} - \eta\bar{\xi}' e^{\phi-2\bar{\phi}-\widetilde{\phi}} + \tfrac{1}{2} cc'\bar{\xi}'\bar{\xi}'' e^{-3\bar{\phi}-\widetilde{\phi}})$$

$$\Rightarrow \quad iY\overline{Y} = \frac{1}{\sqrt{2}} c(\xi' e^{-2\phi-\bar{\phi}} + \bar{\xi}' e^{-\phi-2\bar{\phi}}) e^{-\widetilde{\phi}} \quad ,$$
$$:iX\overline{X}: = :e^{\phi+\bar{\phi}+\widetilde{\phi}}[\beta, Q]: \equiv \left[:\delta(\beta)[\beta,Q]:\right]\left[\delta(\tilde{\beta})\mu\right]$$
$$= e^{\phi+\bar{\phi}+\widetilde{\phi}}\psi\cdot p$$
$$\quad + \left\{i\frac{1}{\sqrt{2}} e^{\bar{\phi}+\widetilde{\phi}}\left[c\xi' + e^{2\phi}\eta' b + (e^{2\phi}\eta b)'\right] + h.c.\right\} \quad ,$$

$$[\beta, Q] = \psi\cdot p - 2\gamma b - c\beta' - \tfrac{3}{2} c'\beta \quad ,$$
$$|OSp\rangle = :X\overline{X}:|0\rangle \quad ,$$
$$|\Phi\rangle = V|0\rangle = V_0|OSp\rangle \quad \Rightarrow \quad V = :V_0 X\overline{X}: \quad . \quad (5.7)$$

The ambiguities in the normal ordering have been fixed by requiring that $:X\overline{X}:$ is the inverse of $Y\overline{Y}$. Thus, $:X\overline{X}:$ can be written as the product of the usual minimal picture-changing operator with a trivial nonminimal picture-changing operator. In practice, it is frequently easier to evaluate $XV$ as $[Q, \xi V]$, etc. Both $V$ and $V_0$ are in the BRST cohomology and have conformal weight zero, as do $X$ and $\overline{X}$, but their ghost numbers differ since $X$ has $J = 1$. Then NS vertex operators take the form

$$V_0 = ce^{-(\phi+\bar{\phi}+\widetilde{\phi})}\Omega \quad ,$$
$$W_0 \equiv \{b_0 + f_0, V_0\} = e^{-(\phi+\bar{\phi}+\widetilde{\phi})}\Omega + c\frac{1}{\sqrt{2}}(i\xi' e^{-2\phi-\bar{\phi}-2\widetilde{\phi}} + h.c.)\Omega \quad ; \quad (5.8a)$$
$$V = -[\widehat{G}, c\Omega] = \gamma\Omega + c\{\psi\cdot p, \Omega\} \quad , \quad W = \{\psi\cdot p, \Omega\} \quad \text{for physical } \Omega,$$
$$\widehat{G} \equiv \psi\cdot p - \gamma b - c'\beta \quad , \quad (5.8b)$$

where $\Omega$ has conformal weight $\frac{1}{2}$. In relation to (5.4), $\Omega = \mathbf{W}|_{\theta=0}$ and $\{\psi\cdot p, \Omega\} = \int d\theta \, \mathbf{W}$. The expression $\widehat{G}$ is not $[\beta, Q]$, but is the OSp(D-1,1|2)-invariant generalization of the light-cone vertex insertion $\psi\cdot p$, which differs by the same factors of 2 as in (5.4). (Similar observations have been made in the covariantized light-cone form of the NSR three-string vertex [14], which differs from the string field theory vertex only by a conformal transformation). (5.4) is thus reproduced, since the $\widehat{G}$ commutator has the same effect as our modified $\theta$ integration. For example, for the massless vector $\Omega = \psi e^{-ik\cdot x}$ reproduces (5.5).

We can extend the prescription for picture changing to integrated vertex operators: Introducing one factor of $X$ at a time,

$$\mathcal{W}_2 \equiv \left\{Q, \int \xi W_1\right\}$$
$$= \int (XW_1 - \xi V_1') = \int (XW_1 + \xi' V_1) \quad ,$$
$$\{Q, XW_1 + \xi' V_1\} = XV_1' + X'V_1 = (XV_1)'$$
$$\Rightarrow \quad V_2 = XV_1 = [Q, \xi V_1] \quad ,$$
$$W_2 = XW_1 + \xi'V_1 = \{Q, \xi W_1\} + (\xi V_1)' \quad . \quad (5.9)$$

The fact that no explicit $\eta$'s, $\xi$'s, or $\phi$'s, or the equivalent $\delta(\beta)$'s or $\delta(\gamma)$'s, appear in $V$, as opposed to $V_0$, is a reflection of the fact that the vacuum $|0\rangle$ is in the same "picture" as the physical states. The exponentials appearing in $V_0$ are those which arise in the path-integral formalism due to $R\phi$-type terms in the classical mechanics action, when the world-sheet curvature $R$ is concentrated at the point of the vertex insertion. In the physical picture, these picture-changing factors just cancel the similar factors in the picture-changing operators $:X\overline{X}:$, leaving the operator $\widehat{G}$ as a vertex insertion. In that sense there is no picture changing in this formalism, and the picture-changing operators do not appear in the expressions actually used: In the operator formalism for conformal field theory, we use expressions such as (5.4) directly; similarly, in the operator expression for the string field theory 3-string vertex it is $\widehat{G}$ that appears and not $:X\overline{X}:$. In the NS sector these expressions are slightly easier to use than the vertex operators of other pictures, since no exponentials or $\delta$-functions occur.

## 5.3. R vertex operators

While the nonminimal coordinates were not needed for the physical NS vertex operators, they are crucial in the Ramond string. The first step in finding R vertex operators is to compare boundary conditions in the NS and R sectors. From the discussion of section 4, looking at physical states in each sector, as well as at the OSp(1|2)-invariant vacuum, we have

| annihilation operators for $n \geq$: | $\|OSp\rangle$ | NS | R |
|---|---|---|---|
| $\frac{1}{\sqrt{2}}(\gamma - i\tilde{\gamma})$ | $\frac{3}{2}$ | $\frac{1}{2}$ | 0 |
| $\frac{1}{\sqrt{2}}(\gamma + i\tilde{\gamma})$ | $\frac{3}{2}$ | $\frac{1}{2}$ | 1 |
| $\frac{1}{\sqrt{2}}(\beta - i\tilde{\beta})$ | $-\frac{1}{2}$ | $\frac{1}{2}$ | 0 |
| $\frac{1}{\sqrt{2}}(\beta + i\tilde{\beta})$ | $-\frac{1}{2}$ | $\frac{1}{2}$ | 1 |
| $\chi$ | $\frac{3}{2}$ | $\frac{1}{2}$ | 0 |
| $\mu$ | $-\frac{1}{2}$ | $\frac{1}{2}$ | 1 |
| $c$ | 2 | 1 | 1 |
| $b$ | $-1$ | 0 | 0 |

where we have chosen the "first" term of R vertex operators: in the language of section 3.3, the "field" ($b_0 = 0$) part before unitarily transforming from the OSp(1,1|2) representation (3.11) back to the usual BRST representation (3.1). For the NS sector, this change in ghost boundary conditions from the OSp(1|2)-invariant vacuum to physical bosonic states is represented by the ghost factors in (5.8a). For the massless spinor, we find

$$|s\rangle = ce^{-(3\phi+\bar{\phi}+3\widetilde{\phi})/2} u^\alpha S_\alpha e^{-ik\cdot x} |OSp\rangle \quad , \qquad (5.10)$$

and using the relation (3.13),

$$V_0 = \frac{\sqrt{2}}{4}[3ce^{-(3\phi+\bar{\phi}+3\widetilde{\phi})/2} - c\bar{\eta}\xi' e^{-(5\phi-\bar{\phi}+3\widetilde{\phi})/2}$$
$$+ i\sqrt{2}cc'\xi' e^{-(5\phi+\bar{\phi}+\widetilde{\phi})/2}] e^{i\pi/4} u^\alpha S_\alpha e^{-ik\cdot x} \quad . \quad (5.11)$$

The final vertex operator is then

$$V = e^{i\pi/4} u^\alpha e^{-ik\cdot x} \left\{ i\frac{1}{\sqrt{2}} ce^{-(\phi-\bar{\phi}-\widetilde{\phi})/2} S_\alpha - \frac{1}{2} ce^{-(\phi-\bar{\phi}+\widetilde{\phi})/2} p_{\alpha\beta} S^\beta \right.$$
$$- \left[ \eta e^{(\phi+\bar{\phi})/2} + \bar{\eta}' e^{(-\phi+3\bar{\phi})/2} - \frac{1}{2} cc'\xi' e^{(-3\phi+\bar{\phi})/2} \right.$$
$$\left. + \frac{2}{3}\bar{\eta} e^{-\phi/2} \left(e^{3\bar{\phi}/2}\right)' + \frac{1}{2} bc\bar{\eta} e^{(-\phi+3\bar{\phi})/2} \right] e^{-\widetilde{\phi}/2} S_\alpha \right\} \quad , \quad (5.12)$$

where $p_{\alpha\beta} \equiv \gamma^a_{\alpha\beta} p_a$.

Interestingly enough there is another picture where the vertex operators are simpler, and also hermitian. As in the other pictures, the vertex operators have conformal weight zero and are in the BRST operator cohomology, but differ in ghost number. This "hermitian" picture is related to the OSp picture and the physical picture by

$$\widehat{V} = XV_0 = \overline{XV}_0 \quad ; \quad V = \overline{X}\widehat{V} \quad , \quad \overline{V} = X\widehat{V} \quad . \quad (5.13)$$

This should be contrasted with the NS vertices, where the "intermediate" vertices are the ones which are nonhermitian. In the hermitian picture the massless spinor vertex is simply

$$\widehat{V} = e^{i\pi/4} ce^{-(\phi+\bar{\phi}+\widetilde{\phi})/2} u^\alpha S_\alpha e^{-ik\cdot x} \quad . \quad (5.14)$$

Unlike the NS case, for the R string this picture is the one where the vertex operators resemble those of the minimal version of the NSR superstring. In fact, if we were to redefine the R string field $\Phi$ by $\Phi \to \overline{Y}(\frac{\pi}{2})\Phi$, where $XY = 1$, all vertex operators would be simple in the OSp picture, but then the field theory action would be like Witten's, with $X_{min} \to \overline{X}X$ and $Y_{min} \to \overline{Y}Y$. Thus, the nonminimal coordinates $\tilde{\gamma}, \tilde{\beta}, \chi, \mu$ have effectively allowed elimination of $Y_{min}$ from the R kinetic term by allowing us to take its "square root."

The reality condition for string fields works a little differently for the nonminimal coordinates. Consistency with conformal field theory (or CPT in the first-quantized theory) requires that complex conjugation correspond to the conformal transformation $\sigma \to \pi - \sigma$ ($z \to -1/z$). Including the Jacobian factors appropriate for the conformal weights, which are just phase factors, this would mean $\Phi[\zeta(\sigma)] = \overline{\Phi}[e^{i\pi d}\zeta(\pi-\sigma)]$. However, this would conflict with the reality condition for the nonminimal coordinates, as described in section 3.3 for the zero-modes: Under complex conjugation, $\gamma_0 \tilde{\beta}_0$ should not change sign, but they would since $\gamma\tilde{\beta}$ has weight $d = 1$. (Only the product of their phase factors matters because of GSO projection.) We therefore introduce an appropriate BRST and conformally invariant "charge conjugation matrix" $\omega$ as

$$\omega(\tilde{\gamma}, \tilde{\beta}, \chi, \mu)\omega^{-1} = -(\tilde{\gamma}, \tilde{\beta}, \chi, \mu) \quad ,$$
$$\omega(\eta, \bar{\eta}, \xi, \bar{\xi}, \phi, \bar{\phi})\omega^{-1} = (\bar{\eta}, \eta, \bar{\xi}, \xi, \bar{\phi}, \phi) \quad , \quad \omega\widetilde{\phi}\omega^{-1} = \widetilde{\phi} + i\pi \quad ,$$
$$\omega Q\omega^{-1} = Q \quad , \quad \omega X\omega^{-1} = \overline{X} \quad , \quad (5.15a)$$

and use it to impose reality conditions on the string fields and vertex operators:

$$\Phi[\zeta(\sigma)] = \omega\overline{\Phi}[e^{i\pi d}\zeta(\pi-\sigma)] \quad ,$$
$$V = \omega\overline{V}\omega^{-1} \quad . \quad (5.15b)$$

$\omega$ is just a "twist" operator which introduces phase factors. Its action in (5.15b) is similar to that of the usual Dirac charge conjugation matrix on spinors and $\gamma$-matrices.

## 6. SUPERSYMMETRY

Just as the integrated, $k=0$, Yang-Mills vertex in the physical picture gives the momentum operator, the integrated, $k=0$, massless spinor vertex in the physical picture gives the generator of supersymmetry [13]. This is because the massless vector and spinor are related by supersymmetry in the same way as $p_0$ and $q$. Another way to understand this is from the Green-Schwarz formalism, where the Yang-Mills field and the massless spinor are related to the gauge fields for derivatives with respect to the Green-Schwarz coordinates $x^a$ and $\Theta^\alpha$ [15]. Thus, the supersymmetry transformations are given by

$$\delta\Phi = \epsilon^\alpha q_\alpha \Psi \quad, \quad \delta\Psi = \epsilon^\alpha q_\alpha \Phi \quad,$$

$$q_\alpha = \omega \hat{q}_\alpha = \hat{q}_\alpha^\dagger \omega^{-1} \quad, \tag{6.1}$$

for R field $\Phi$ and NS field $\Psi$, where $\hat{q}_\alpha$ is taken from the physical-picture integrated vertex operator $\mathcal{W}$ for the massless spinor as $\mathcal{W}|_{k=0} = u^\alpha \hat{q}_\alpha$. $\omega$ allows the supersymmetry transformation to be written in a more symmetric form: On the space $\binom{\Psi}{\Phi}$, where the supersymmetry generator has the form $\binom{0\ A^\dagger}{A\ 0}$, we have $A = A^\dagger = q$.

To evaluate the closure of the supersymmetry algebra, we write

$$\begin{aligned}
\hat{q} &= \left[Q, \int \xi \widehat{W}\right] \\
&= \left[Q, \xi(\tfrac{\pi}{2}) \int \widehat{W}\right] + \left[Q, \int \{\xi - \xi(\tfrac{\pi}{2})\}\widehat{W}\right] \\
&= X(\tfrac{\pi}{2})\widehat{\mathcal{W}} + \left[Q, \int \{\xi - \xi(\tfrac{\pi}{2})\}\widehat{W}\right] \quad . 
\end{aligned} \tag{6.2}$$

The second term is a true BRST variation, since $\xi_0$ drops out. (In [2] this manipulation was performed for $p_0$. However, the same manipulation can be performed for one of the supersymmetry transformations given there for the minimal theory. Thus, supersymmetry does not mix levels there either, even though the R kinetic term in the minimal theory does.) We therefore have

$$\begin{aligned}
\{q_\alpha, q_\beta\} &= \hat{q}_{(\alpha}^\dagger \hat{q}_{\beta)} \\
&\approx \overline{X}(\tfrac{\pi}{2}) X(\tfrac{\pi}{2})\{\widehat{\mathcal{W}}_\alpha, \widehat{\mathcal{W}}_\beta\} \\
&= \overline{X}(\tfrac{\pi}{2}) X(\tfrac{\pi}{2}) \tilde{p}_{\alpha\beta} \\
&\approx p_{0\alpha\beta} \quad ,
\end{aligned} \tag{6.3}$$

where "$\approx$" means up to $Q$-commutator terms, $\tilde{p} \equiv \int e^{-(\phi+\bar{\phi}+\widetilde{\phi})}\psi$ is $\mathcal{W}$ for the $k=0$ Yang-Mills field in the OSp picture, and we have used the fact that $X$ commutes with $\widehat{\mathcal{W}}$. The $Q$-commutator terms give two types of terms: When acting on a field, $\{Q, A\}\Phi = QA\Phi + AQ\Phi$, the first term is a gauge transformation and the second term is proportional to the field equations (the trivial gauge symmetry of the form $\delta\phi_i = \epsilon_{ij}\delta S/\delta\phi_j$ mentioned in section 2.1).

Such trivial non-closure terms are common in component formulations of supersymmetric theories. If this theory had been derived from a manifestly supersymmetric formalism (such as a covariantly quantized Green-Schwarz string), the first term would have resulted from going to a Wess-Zumino gauge by eliminating nonderivative-gauge degrees of freedom (similar to $A_+$ or $\kappa$ of section 3.4), and the second term would have resulted from eliminating auxiliary fields (similar to $A_-$ in section 3.4) by their algebraic equations of motion.

For this closure to be maintained at the interacting level (see (2.2a,2.4c)), the interacting contributions to these two terms must cancel [2]:

$$\begin{aligned}
\{Q, A\}\Phi &= QA\Phi + AQ\Phi \\
&= [QA\Phi + \Phi \star (A\Phi) - (A\Phi) \star \Phi] + A(Q\Phi + \Phi \star \Phi)
\end{aligned}$$

$$\Rightarrow \quad A(\Phi \star \Phi) = (A\Phi) \star \Phi - \Phi \star (A\Phi) \quad, \tag{6.4}$$

which means that the operator $A$ must be distributive over the $\star$ product, just as $Q$ is.

## 7. AMPLITUDES

As an example of a conformal field theory calculation, consider the three-Yang-Mills vertex:

$$\langle 0|V_a(k_1, z_1) V_b(k_2, z_2) V_b(k_3, z_3)|0\rangle$$

$$= \int d\theta_1\, d\theta_2\, d\theta_3\, \langle 0|\mathbf{C}(\mathbf{Z}_1)\mathbf{C}(\mathbf{Z}_2)\mathbf{C}(\mathbf{Z}_3)|0\rangle$$
$$\times {}_x\langle 0|\mathbf{W}_a(k_1, \mathbf{Z}_1)\mathbf{W}_b(k_2, \mathbf{Z}_2)\mathbf{W}_c(k_3, \mathbf{Z}_3)|0\rangle_x \quad, \tag{7.1}$$

where $|0\rangle_x$ is the vacuum for just the physical coordinates $\mathbf{X}$.

The $\langle 0|\mathbf{CCC}|0\rangle$ matrix element can easily be evaluated, since the only nonvanishing contributions come from

$$\langle 0|\gamma_{1/2} c_0 \gamma_{-1/2}|0\rangle = 1 \quad, \tag{7.2}$$

since $\gamma_{-1/2}|0\rangle$ is the tachyon. From the usual expansion for a field of conformal weight $d$, $\zeta = \sum \zeta_n z^{-n-d}$, we then have

$$\langle 0|\mathbf{C}(\mathbf{Z}_1)\mathbf{C}(\mathbf{Z}_2)\mathbf{C}(\mathbf{Z}_3)|0\rangle = \theta_1 \theta_2 z_3(z_1 + z_2) + \text{cyclic permutations}. \tag{7.3}$$

We now consider the generalization to the N-point amplitude. We need to include contributions from the string propagators: The propagator is $(b_0 + f_0)/\mathbf{L}_0$, as for the Veneziano string, except that for the Veneziano string $f_0$ is usually taken to be 0. For physical NS vertex operators the $f_0$ won't contribute anyway because of the absence of nonminimal coordinates in these operators. The calculation is similar to the Veneziano case [16]: The separation of the vertices in the complex plane, and the integration over that separation, accounts for the $1/\mathbf{L}_0$'s. The factors of $b_0 + f_0$ appear as contour integrals of $b + f$ surrounding certain sets of vertices, corresponding to those on either side of that propagator in the string field theory Feynman diagram (those inside the contour and those outside). Since these contours can be expressed as sums of contours around individual vertices, we can consider without loss of generality a complex plane with such contours surrounding each of $N - 3$ of the $N$ vertex insertions. (We use the fact that $b_0|0\rangle = f_0|0\rangle = 0$.) Each contour integral gives simply the commutator $\{b_0 + f_0, V\} = W$ (see (5.2)). (Note that $b_0 + f_0$ and $\mathbf{L}_0$ are both integrals over $\rho = \tau + i\sigma$, but the result can be converted to the $z$-plane since $d\rho\, W(\rho) \to dz\, W(z)$ and $V(\rho) \to V(z)$.) These steps are also clear by direct manipulation in the operator formalism. We thus have the amplitude

$$A = \int d\rho_4 \cdots d\rho_N \, \langle 0|VVV(b_0+f_0)V \cdots (b_0+f_0)V|0\rangle$$
$$= \int dz_4 \cdots dz_N \, \langle 0|VVVW \cdots W|0\rangle$$
$$= \int d^N\theta \, dz_4 \cdots dz_N \, \langle 0|\mathbf{C}_1\mathbf{C}_2\mathbf{C}_3|0\rangle \cdot {}_x\langle 0|\mathbf{W}_1 \cdots \mathbf{W}_N|0\rangle_x \ . \quad (7.4)$$

All steps except the last (i.e., substitution in terms of $\int d\theta$ using (5.4)) are the same as for the Veneziano model. (A similar, but less simple evaluation occurs for the particle (3.20).) This calculation applies to arbitrary physical NS vertices. (Of course, unphysical states, including ghost and auxiliary states, involve more general expressions, including dependence on the nonmimimal fields.)

The $F = {}_x\langle 0|\mathbf{W} \cdots \mathbf{W}|0\rangle_x$ matrix element is the same as that obtained from the "old covariant" approach [17]. There the amplitude was written as

$$A_{oc} = (z_1 - z_3)(z_3 - z_2) \int d\theta_3\, d\theta_4 \cdots d\theta_N \, d^{N-3}z \, F \ . \quad (7.5)$$

This amplitude is Sp(2) invariant, and invariant in particular under the cyclic permutation of the first three $\mathbf{Z}$'s. If we plug (7.3) into (7.4), noting that the two $\theta$'s just kill two of the $\theta$ integrals, we find

$$A = \left(\frac{z_3(z_1+z_2)}{(z_1-z_3)(z_3-z_2)} + \text{cyclic permutations}\right) A_{oc} = A_{oc} \ , \quad (7.6)$$

reproducing the old result.

Except for the manifest world-sheet supersymmetry (which holds only for the NS sector anyway), this calculation is similar to that using the minimal coordinates [13]. Because there the OSp vacuum was used, two $V_0$ vertices were needed, the rest were $V$'s (see section 5.2; this is easily derived using the minimal analog of (5.7)). However, since the OSp vacuum satisfies $\langle OSp|ccc|OSp\rangle \sim 1$ instead of $\langle 0|\gamma\gamma c|0\rangle \sim 1$ we obtain the same result, since the $c$ term (the only term) of $V_0$ is essentially the same as the $\gamma$ term of $V$.

Using the techniques described in section 5, R vertex operators can be constructed and used to calculate amplitudes. The explicit expressions for the R vertex operators in the physical picture are messier than in the usual, minimal formalism. (See e.g., (5.12).) However, many of the terms do not contribute, since there must be a balance in the numbers of each of the different kinds of ghost factors. On the other hand, calculations can also be performed in mixed pictures, as in the usual approach, with simple expressions such as (5.14). It is clear that working with our new nonminimal coordinates does not offer any computational advantages to the usual approach when applied to conformal field theory. The main advantage in the conformal field theory is conceptual: All states and all vertex operators can have the same ghost number, so quantization of the NSR string requires no concepts that weren't already understood from the Veneziano string. (This could already be done for the NS sector even in the minimal formalism.) The reason that this does not simplify the treatment of the R sector is that the conformal field theory approach requires that states and vertex operators be expressed with respect to the NS vacuum. On the other hand, in the string field theory approach the NS and R strings can have their own separate vacuua, avoiding this problem, so the treatment of the R string field is significantly simpler than in the usual, minimal treatment.

## 8. NSR SUPERSTRING FIELD THEORY

In the string field theory, the separate treatment of the NS and R strings, together with the use of the same "picture" (i.e., ghost number) for these two strings, allows the avoidance of both bosonization and $\delta$-functions of ghosts. This means that vertex operators have simple forms, such as (5.4,5) for the NS string, and (3.13) for the massless spinor. In the old formulation bosonization and ghost $\delta$-functions were avoidable in some terms in the action, but not for the R kinetic term.

As described above, the gauge invariant kinetic terms for both the NS and R string fields are just $\langle \Phi|Q\Phi \rangle$, as in string field theory for the Veneziano string. This agrees with Witten's original version of NSR superstring field theory [2] in the NS sector, except that $Q$ now has the nonminimal term. Of course, this new $Q$ is the same one in both the NS and R sectors, except that the NS and R fields satisfy different boundary conditions (or, equivalently, the mode expansion of $Q$ differs).

The expression for the $(NS)^3$ vertex is also the same as in Witten's version up to nonminimal stuff, since the NS sector never really had a picture-changing problem. (The apparent problem in the conformal field theory was due to the choice of a vacuum which did not correspond to any state in the string field theory.) This means that there is still an operator insertion at the vertex, but it is not related to ghost-number problems: In fact, this vertex insertion, although called the "picture-changing operator," has dependence on non-ghost coordinates, in contrast to the so-called "inverse picture-changing operator" that appeared in the old versions of the R-string kinetic operator [18,2], which depended only on ghosts. We therefore have a situation similar to light-cone string field theory [19], where the kinetic operators are both $\mathbf{L}_0$, and both vertices have a vertex insertion, similar to the one here.

However, because the R-string kinetic operator is now the same as for the NS string, the two vertices, $(NS)^3$ and $NS(R)^2$, are also the same. We can therefore effectively write the NS and R string fields collectively as a single field. Also, unlike previous treatments of NSR superstring field theory, the same operator performs supersymmetry transformations on the NS and R string fields, correctly changing the boundary conditions in both cases.

The explicit expressions for these two vertices appearing in the action (2.2) are therefore similar to those used for the $(NS)^3$ vertex in the minimal case [20-22]. Each vertex is defined as the product of the "naive" one with the factor $iX\overline{X}$, inserted at the interaction point (midpoint). The "naive" one is that obtained e.g., by functionally integrating $e^S$ over the infinitesimal strip connecting the 3 strings, and thus includes not only $\delta$-functionals of the coordinates but also $e^\phi$-type factors coming from the $R\phi$ terms in the first-quantized action for the bosonized ghosts. In terms of analytic variables, there are effectively two interaction points ($\sigma = \pi/2$ or $-\pi/2$), so the interaction vertex can be defined to have $X$ inserted at either point, and the same for $\overline{X}$ (so $X$ and $\overline{X}$ need not be at the same point). In practice the simplest method of analysis is that of Suehiro [22]: The 2D Green functions for the physical variables and unbosonized ghosts, which define the vertex operator, are evaluated with appropriate boundary conditions at both the positions $z_r$ of the external strings and the interaction point $z_0$ (and its mirror image $\bar{z}_0$). This means that the $\rho$-plane Green functions are expressed in terms of $z$ as just $1/(z-z')$ times appropriate powers of $z-z_0$, $z-\bar{z}_0$, and $z-z_r$ (and of similar factors with $z'$). In particular, the boundary conditions at $z_0$ are chosen to absorb the $e^\phi$ factors from $X$ and $\overline{X}$, so the remaining insertion at the interaction point is an expression of the form $\psi \cdot p + \cdots$, similar to $\widehat{G}$ of (5.8b), which can be expressed directly in terms of the unbosonized ghosts. This remaining insertion is also expressed in terms of the Green functions. The mode expansion of the Green functions is then performed as by Gross and Jevicki [20].

It is also interesting to analyze vertices which correspond to endpoint interactions, instead of midpoint ones. Although such interactions do not give the correct field theory, they are equal to the correct midpoint ones on the (free) mass shell. Such interactions occur on the real axis (the string boundary) in the $z$-plane, $z_0 = \bar{z}_0$, so there is a simplification in both the Green function boundary conditions and the vertex insertion. There is also a simplification in interpretation, since the Jacobian factors $\partial\rho/\partial z \sim \sqrt{(z-z_0)(z-\bar{z}_0)}/\prod(z-z_r)$ [16] simplify to $(z-z_0)/\prod(z-z_r)$. The result is that the ghosts can be considered to have the same conformal weights as the corresponding physical fields ($c$ and $b$ the same as $x$ and $p$; $\beta$, $\gamma$, $\tilde{\beta}$, $\tilde{\gamma}$, $\chi$, and $\mu$ the same as $\psi$), leaving a vertex insertion of $c\widehat{G}$. The result is covariant with respect to $OSp(10,2|4)$ (or $OSp(9,1|2)$ in the minimal case [14]), which should have been expected since the endpoint interaction can be derived directly by adding extra dimensions to the light-cone result [7], which also uses endpoint interactions.

Just as the interaction term can be expressed as $\langle V_3|\Phi\rangle|\Phi\rangle|\Phi\rangle$, the supersymmetry transformation can be expressed as $\langle q|\Phi\rangle|\Phi\rangle$, where one $\Phi$ is an R string and one an NS string. This $\langle q|$ actually represents the operator $\hat{q}$ of (6.1), and its mode expansion can be calculated in the same manner as the 3-string vertices, without bosonization or ghost $\delta$-functions. The $\omega$ factor which appeared in $q$ in (6.1) arises from the reality condition: The reality condition on the string field can be expressed as $\langle\Phi|\Phi'\rangle = \langle I_2|\Phi\rangle|\Phi'\rangle$ [23], where $\langle I_2|$ has an additional factor of $\omega$ compared to the naive value. This is because the naive expression is just the usual twist operator, expressed functionally as $\delta[\zeta_1(\sigma) - e^{i\pi d}\zeta_2(\pi-\sigma)]$ in terms of $\delta$-functionals of coordinates $\zeta$ of conformal weight $d$ between strings 1 and 2. However, all the nonminimal coordinates have been assigned conformal weights $\frac{1}{2} \pm 1$, whereas their natural weights are $\frac{1}{2}$. (A conformal weight of $\frac{1}{2}$ requires no $e^\phi$ insertions, since there are then no $R\phi$ terms in the first-quantized action.) Equivalently, the OSp formalism discussed above, which holds for the kinetic terms in all cases, gives the nonminimal coordinates the same weight $\frac{1}{2}$ as for $\psi$. The $\omega$ factor compensates for this by changing the relative sign of $\zeta_1$ and $\zeta_2$ for the nonminimal coordinates, fixing the $e^{i\pi d}$ factor.

Because the $(NS)^3$ and $NS(R)^2$ interaction terms now both require an operator insertion, the usual contact-term divergences [24] are present in all four(or more)-string NSR scattering amplitudes. This situation is similar to that of the non-supersheet light-cone formalism, where operator insertions are also necessary for both $(NS)^3$ and $NS(R)^2$ interactions. Other authors [25] have tried to eliminate these contact-term divergences in the string field theory by considering kinetic terms of the form $\langle\Phi|ZQ\,\Phi\rangle$, where $Z$ is a BRST-invariant operator inserted at the midpoint. However, if one restricts $\Phi$ to satisfy the usual boundary conditions (i.e., it is constructed out of the ground state with a finite set of oscillator modes), it is not possible to gauge fix these types of kinetic terms to the simple form $\mathbf{L}_0$ [26]. To be more specific, gauge transformations of the type $\Phi \rightarrow \Phi + B$ for $ZB = 0$ are not allowed since $B$ cannot be constructed out of the ground state with a finite set of oscillator modes. Without the use of these gauge transformations, it is impossible to define a propagator for such formulations which avoids the divergences in the four-string amplitudes off shell. On the other hand, if one were to allow arbitrary boundary conditions for $\Phi$, the $\star$ product would lose its associativity, making the interaction term ill-defined. At this point, it is unclear if some "compromise" set of boundary conditions can be found which would allow the gauge transformations necessary for such a formalism, but still preserve the associativity property of the $\star$ product.

## 9. CONCLUSIONS

In this paper we have given a new formulation of NSR superstring theory which avoids picture changing. The main difference is the use of nonminimal fields which solve the bosonic zero-mode problem of the R sector, allowing the construction of the Hilbert space and BRST formalism in a way more similar to that of the Veneziano string. In particular, in the string field theory the Ramond string is then treated in exactly the same way as the Neveu-Schwarz string except for boundary conditions. In the conformal field theory, we choose as the new vacuum the unique string field theory state in the BRST cohomology with ghost number $-\frac{3}{2}$. As a result, vertex operators are unique, as opposed to the old approach, where picture changing was required even for the NS sector.

It might be interesting to see if such methods would suggest a method for obtaining a covariant Green-Schwarz formulation. Since the GS and NSR formalisms are directly related in the light cone by triality, a covariant form of triality might appear if an appropriate set of ghosts (that which closes supersymmetry off shell) is chosen. With regard to the OSp analysis of section 3, it should be noted that the set of auxiliary scalars agreeing with those obtained by superfield methods for N=1 supersymmetry in various dimensions is given by OSp(2,1|2) in D=3, OSp(4,2|4) in D=4, and OSp(8,4|8) in D=6, suggesting that perhaps OSp(16,8|16) might be appropriate for D=10 and thus for superstrings. On the other hand, the SO(D,2) "physical" submultiplets occurring in the present formalism are strongly suggestive of broken conformal invariance, as commonly used in supergravity.

## ACKNOWLEDGMENT

N.B. thanks Christian Preitschopf for useful discussions. M.T.H. thanks Kareljan Schoutens for conversations on normal ordering. W.S. thanks Peter van Nieuwenhuizen for discussions on ZJBV quantization and for pointing out the first reference of [5], and Jon Yamron for suggesting the title.

## REFERENCES

[1] W. Siegel, *Int. J. Mod. Phys. A* **6** (1991) 3997.
[2] E. Witten, *Nucl. Phys.* **B268** (1986) 253, **276** (1986) 291.
[3] E. Witten, Some remarks about string field theory, *in* Marstrand Nobel Sympos. 1986, p. 70;
I.B. Frenkel, H. Garland, and G.J. Zuckerman, *Proc. Natl. Acad. Sci. USA* **83** (1986) 8442.
[4] W. Siegel, Introduction to string field theory (World Scientific, Singapore, 1988) p. 151.
[5] J. Zinn-Justin, *in* Trends in elementary particle theory, eds. H. Rollnik and K. Dietz (Springer-Verlag, Berlin, 1975);
I.A. Batalin and G.A. Vilkovisky, *Phys. Lett.* **102B** (1983) 27, **120B** (1983) 166; *Phys. Rev.* **D28** (1983) 2567, **D30** (1984) 508; *Nucl. Phys.* **B234** (1984) 106; *J. Math. Phys.* **26** (1985) 172.
[6] W. Siegel, *Int. J. Mod. Phys. A* **4** (1989) 3705.
[7] W. Siegel and B. Zwiebach, *Nucl. Phys.* **B282** (1987) 125, **299** (1988) 206;
Ref. 4, pp. 38, 216.
[8] W. Siegel, *Nucl. Phys.* **B284** (1987) 632;
W. Siegel, Int. J. Mod. Phys. **A4** (1989) 1827;
Universal supersymmetry by adding 4+4 dimensions to the light cone, *in* Strings '88, eds. S.J. Gates, Jr., C.R. Preitschopf, and W. Siegel, College Park, MD, May 24-28, 1988 (World Scientific, Singapore, 1989), p. 110;
Ref. 4, pp. 45, 46, 60, 75, 110, 168, 211.
[9] Ref. 4, pp. 27, 81, 101, 120.
[10] L. Baulieu, W. Siegel, and B. Zwiebach, *Nucl. Phys.* **B287** (1987) 93.
[11] W. Siegel, Lorentz-covariant gauges for Green-Schwarz superstrings, *in* Strings '89, eds. R. Arnowitt, R. Bryan, M.J. Duff, D. Nanopoulos, and C.N. Pope (World Scientific, Singapore, 1989) p. 211.
[12] S.J. Gates, Jr., M.T. Grisaru, U. Lindström, M. Roček, W. Siegel, P. van Nieuwenhuizen, and A.E. van de Ven, *Phys. Lett.* **225B** (1989) 44;
U. Lindström, M. Roček, W. Siegel, P. van Nieuwenhuizen, and A.E. van de Ven, *Nucl. Phys.* **B330** (1990) 19;
M.B. Green and C.H. Hull, *Phys. Lett.* **225B** (1989) 57;
R. Kallosh, *Phys. Lett.* **224B** (1989) 273.
[13] D. Friedan, E. Martinec, and S. Shenker, *Phys. Lett.* **160B** (1985) 55; *Nucl. Phys.* **B271** (1986) 93;
V.G. Knizhnik, *Phys. Lett.* **160B** (1985) 403;
V.A. Kostelecký, O. Lechtenfeld, W. Lerche, S. Samuel, and S. Watamura, *Nucl. Phys.* **B288** (1987) 173.
[14] H. Hata, K. Itoh, T. Kugo, H. Kunitomo, and K. Ogawa, *Prog. Theor. Phys.* **78** (1987) 453.
[15] W. Siegel, *Nucl. Phys.* **B263** (1986) 93.
[16] S.B. Giddings, *Nucl. Phys.* **B278** (1986) 242;
S.B. Giddings and E. Martinec, *Nucl. Phys.* **B278** (1986) 91.
[17] L. Brink and J.-O. Winnberg, *Nucl. Phys.* **B103** (1976) 445.
[18] J.P. Yamron, *Phys. Lett.* **174B** (1986) 69.
[19] S. Mandelstam, *Nucl. Phys.* **B69** (1974) 77.
[20] D. Gross and A. Jevicki, *Nucl. Phys.* **B293** (1987) 29.
[21] S. Samuel, *Nucl. Phys.* **B296** (1988) 187.
[22] K. Suehiro, *Nucl. Phys.* **B296** (1988) 333.
[23] A. LeClair, M. Peskin, and C.R. Preitschopf, *Nucl. Phys.* **B317** (1989) 411.
[24] J. Greensite and F.R. Klinkhamer, *Nucl. Phys.* **B281** (1987) 269, **291** (1987) 557;
M.B. Green and N. Seiberg, *Nucl. Phys.* **B299** (1988) 559;
C. Wendt, *Nucl. Phys.* **B314** (1989) 209.
[25] S.P. Martin, *Nucl. Phys.* **B310** (1988) 428;
C.B. Thorn, *Phys. Rep.* **175** (1989) 1;
C.R. Preitschopf, C.B. Thorn, and S. Yost, *Nucl. Phys.* **B337** (1990) 363.

[26] I. Ya. Aref'eva and P.B. Medvedev, *Phys. Lett.* **202B** (1988) 510;
O. Lechtenfeld and S. Samuel, *Phys. Lett.* **213B** (1988) 431;
R. Bluhm and S. Samuel, *Nucl. Phys.* **B338** (1990) 38;
I. Ya. Aref'eva, P.B. Medvedev, and A.P. Zubarev, *Nucl. Phys.* **B341** (1990) 464.